# Impact of eHMI on Pedestrians' Interactions with Level-5 Automated Driving Systems

Viktoria Marcus[1], Griffin Pitts[2], Sanaz Motamedi[3]

1 Industrial and Systems Engineering, University of Florida
2 Computer and Information Science and Engineering, University of Florida
3 Industrial and Manufacturing Engineering, Pennsylvania State University

Each year, over half of global traffic fatalities involve vulnerable road users (e.g. pedestrians), often due to human error. Level-5 automated driving systems (ADSs) could reduce driver errors contributing to pedestrian accidents, though effectiveness depends on clarity and understandability for other road users. External human-machine interfaces (eHMIs) have been proposed to facilitate pedestrian-ADS communication, though consensus on optimal eHMI features remains unclear. In an online survey, 153 participants responded to road-crossing scenarios involving level-5 ADSs, with and without eHMIs. With eHMIs, pedestrians crossed earlier and more confidently, and reported significantly increased perceptions of safety, trust, and understanding when interacting with level-5 ADSs. Visual eHMI features (including a text display and external speedometer) were ranked more necessary than auditory ones, though auditory cues received positive feedback. This study demonstrates that eHMIs can significantly improve pedestrians' understanding of level-5 ADS intent and enhance perceived safety and trust, facilitating more intuitive pedestrian-ADS interactions.

## INTRODUCTION

According to the World Health Organization (WHO), over half of the 1.19 million annual traffic fatalities involve vulnerable road users (WHO, 2023). Vulnerable road users are those who are at higher risk of injury or death in a traffic accident, especially pedestrians (WHO, 2023). Human error remains a major factor in traffic accidents (Hafeez et al., 2023), suggesting that level-5 Automated Driving Systems (ADSs) could significantly improve safety. These vehicles, capable of "sustained and unconditional" operation without human supervision (Shi et al., 2020), could eliminate many human-caused errors. However, their safety benefits depend on their capability to effectively interact with other road users, particularly pedestrians, an area that remains poorly understood (Ferenchak & Shafique, 2022; Habibovic et al., 2018; Jayaraman et al., 2020).

Pedestrians typically rely on implicit cues (e.g. speed) and explicit signals (e.g. eye contact) to assess right-of-way (Habibovic et al., 2018; Matthews et al., 2017; Wang et al., 2021; Zhao et al., 2022). While implicit cues remain available with level-5 ADSs, the lack of a human driver disrupts explicit communication. To bridge this gap, external human-machine interfaces (eHMIs), which are visual and auditory displays, have been proposed to convey vehicle intent (Ferenchak & Shafique, 2022; Habibovic et al., 2018; Métayer & Coeugnet, 2021; Rasouli & Tsotsos, 2020; Wang et al., 2021). Although eHMIs generally improve interactions, there is no consensus on which features are most helpful (Deb et al., 2018; Dey et al., 2021; Ferenchak & Shafique, 2022; Matthews et al., 2017).

This study examines how an eHMI designed by Marcus et al. (2024) affects pedestrians' crossing decisions and identifies preferred features through a survey using text and image-based interaction scenarios.

## LITERATURE REVIEW

**Pedestrians in Traffic**

In regular traffic, pedestrians often use implicit cues to gauge driver intent (e.g. changes in speed, gap size, traffic flow). Then, as a vehicle approaches a pedestrian, they may confirm driver intent through explicit signals (e.g. eye contact, hand gestures) for confirmation (Habibovic et al., 2018; Matthews et al., 2017; Wang et al., 2021; Zhao et al., 2022). This interaction reassures them of being seen and clarifies intent (Jayaraman et al., 2020; Rasouli & Tsotsos, 2020). In level-5 ADSs, the absence of a human driver to communicate with leaves a gap in interactions between them and pedestrians.

Studies of pedestrian behavior around ADSs show mixed results: some report riskier behavior, assuming ADSs will always yield (Zhao et al., 2022), while others observe greater caution (Habibovic et al., 2018). Although eHMIs have been widely studied, no consensus exists on optimal design. However, eHMIs generally enhance pedestrians' willingness to cross, trust in ADSs, situational awareness, and overall interaction quality (Deb et al., 2018; Dey et al., 2021; Ferenchak & Shafique, 2022; Matthews et al., 2017).

**eHMI Design**

Despite agreement on their value, design elements of eHMIs remain debated. Recommendations include signaling automation, conveying vehicle intent (rather than commands), confirming pedestrian detection, and using multimodal signals in strategic locations (Habibovic et al., 2018; Rasouli & Tsotsos, 2020; Lagström & Lundgren, 2015; Wang et al., 2021; Eisma et al., 2019). Tested features include LED lights, symbols, text displays, and audio cues like speech and chimes (Deb et al., 2018; Dey et al., 2021; Matthews et al., 2017; Métayer & Coeugnet, 2021). Though generally useful, specific guidelines remain unresolved.

Marcus et al. (2024) applied a co-design approach with participants to understand how to create eHMI elements that facilitated safe pedestrian-ADS interactions. Elements included speed indicators, timers, lights, text, labels, and audio cues designed to enhance safety and trust, clarify the ADS's intent, and convey pedestrian detection. This study evaluates an eHMI inspired by their findings via a text-based scenario survey, asking: (1) How does such an eHMI affect crossing decisions? and (2) Which eHMI features do pedestrians consider essential?

## METHODOLOGY

**eHMI Development**

The eHMI included the features that were most often included in designs from Marcus et al. (2024), including an external speedometer, pedestrian silhouette, text display, automated vehicle label, and audio speaker (Figure 1).

**Survey**

To address the research questions, an online survey was developed using Qualtrics. The survey included an IRB approved informed consent and background information section, then introduced the concept of a level-5 ADS through 2-minute video. Next, participants responded to a text-based road-crossing scenario involving a level-5 ADS without an eHMI. Then, the survey introduced the concept of an eHMI in a paragraph and presented the eHMI design based on Marcus et al. (2024).

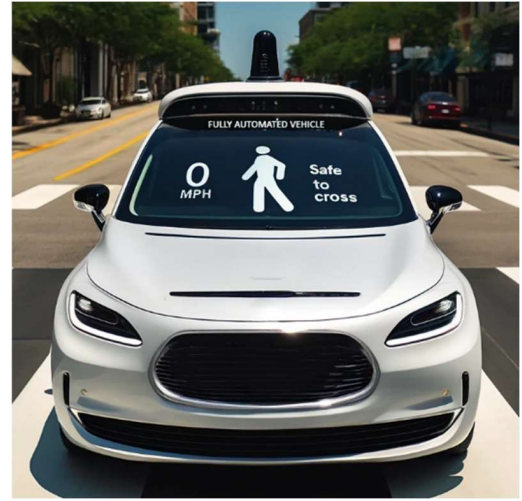

Figure 1. The eHMI used in this study.

Subsequently, participants evaluated a second scenario identical to the first, but with the level-5 ADS equipped with the described eHMI. Finally, participants were asked to rank and provide feedback on the different features of the eHMI.

The text-based scenarios asked participants to select the option most similar to how they would react to crossing an unsignalized zebra crosswalk in front of an approaching level-5 ADS, without (Scenario 1) and with (Scenario 2) an eHMI. To develop the scenarios and response options, previous studies were used as inspiration, including those by Zhao et al. (2022) and Deb et al. (2018). The response options gave a range of risk levels, from crossing as soon as the ADS is seen, to avoiding crossing altogether. A write-in option was also provided. To evaluate the eHMI, participants were asked to rank eHMI features in a matrix table using a five-point Likert scale, as well as reflect on its impacts on their perceptions of safety, trust, and understanding. Optionally, participants could provide written feedback on the overall design.

**Participants**

The survey was shared through researcher contacts, social media, and professional organization discussion boards. After filtering for completeness and removing responses that showed more than a two-point difference on validation questions, 153 responses remained. The participant sample had an average age of 28.14 (SD=12.23). Nearly half of the participants (49.67%) identified as male, 46.41% female, and 3.92% identified as non-binary, other, or preferred to self-describe their gender.

## RESULTS

**Road Crossing Scenarios**

To understand the impact of an eHMI on pedestrians' road-crossing decisions in front of a level-5 ADS, the responses between scenarios without-eHMI (Scenario 1) and with-eHMI (Scenario 2) were compared. Chi-square test of independence was conducted to examine the relationship between the presence of eHMI and pedestrian crossing behavior. The test revealed a significant association ($\chi^2$ (6, N = 153) = 15.53, p = 0.017), indicating that the presence of eHMI influenced pedestrian crossing decisions.

In the scenario without eHMI (Scenario 1), the most common behavior was to "Wait for ADS to stop, then cross in front" (41.18%), followed by "Cross when sure ADS detects you" (21.57%) and "Cross when ADS slows down" (21.57%). Only 0.65% of pedestrians chose to cross immediately, while 7.19% opted not to cross at all, suggesting a general caution when interacting with ADSs without eHMI. The remaining 2.61% wrote in their answer, usually opting not to cross or waiting to confirm that the ADS was not just an uncontrolled vehicle. This distribution is shown in Table 1.

The introduction of the eHMI (Scenario 2) led to significant shifts in pedestrian behavior, generally indicating increased confidence in crossing. The proportion of pedestrians choosing to "cross immediately" increased from 0.65% to 8.50%. Simultaneously, there was a decrease in more cautious behaviors. The percentage of pedestrians who waited for the ADS to stop before crossing in front dropped from 41.18% to 30.07%, and the same percentage (21.57%) waited for the level-5 ADS to slow before crossing. There was an increase in pedestrians crossing when sure of ADS detection, rising from 21.57% to 28.10%, which indicates that pedestrians were able to confirm the yielding intention of the ADS using the eHMI. In addition, fewer pedestrians (3.27%) crossed behind the ADS after stopping with the eHMI than without (5.23%). The percentage of pedestrians who chose not to cross at all dropped from 7.19% to 5.23%. Those who wrote in their answer usually said they would wait for an indication from the eHMI, start crossing immediately but not with full confidence, or not cross at all. These results are summarized in Table 1.

**Impact of eHMI**

Respondents were asked to compare levels of safety, trust, and ease of understanding the level-5 ADS's intent when it was equipped with the eHMI versus without. The distributions are shown in Table 2. Regarding safety, 79.74% of participants reported that they felt safer around the level-5 ADS when it was equipped with the eHMI, 5.23% felt the same level of safety, and 15.03% felt less safe. When crossing the road in front of the level-5 ADS, 75.16% of pedestrians had more trust in the vehicle when it was equipped with the eHMI versus without, 5.23% had about the same level of trust, and 19.61% had less trust. When the level-5 ADS was equipped with the eHMI, 84.31% of pedestrians found it easier to understand the intention of the vehicle, 3.27% felt about the same, and 12.42% found it more difficult to understand the level-5 ADS's intent.

Pedestrians were also asked to consider if they would prefer a level-5 ADS to have an eHMI or not when crossing the road in front of one in the future. A majority (87.58%) of participants responded that they would prefer the level-5 ADS to have an eHMI, while 2.61% did not have a preference, and 9.80% preferred not to have it.

Table 1. Distribution of responses in road crossing scenarios without (1) and with (2) the eHMI.

| Response | Scenario 1 | Scenario 2 |
|---|---|---|
| Cross immediately | 0.65% | 8.50% |
| Cross when ADS slows down | 21.57% | 21.57% |
| Cross when ADS detects you | 21.57% | 28.10% |
| Wait for ADS to stop, cross in front | 41.18% | 30.07% |
| Wait for ADS to stop, cross behind | 5.23% | 3.27% |
| Do not cross | 7.19% | 5.23% |
| Other (write in) | 2.61% | 3.27% |

Table 2. Distribution of perceptions of safety, trust, and understanding with the eHMI.

| | More | About the Same | Less |
|---|---|---|---|
| Safety | 79.74% | 5.23% | 15.03% |
| Trust | 75.16% | 5.23% | 19.61% |
| Understanding | 84.31% | 3.27% | 12.42% |

**eHMI Features**

When ranking which eHMI features were most useful, participants ranked the features in the following order, from most to least necessary: text display, vehicle movement, automated vehicle label, external speedometer, pedestrian silhouette, and audio speaker. This suggests that participants placed more value on visual information compared to auditory cues. The high rankings of the text display and external speedometer aligns with previous studies which emphasized the importance of explicit communication in eHMI design (Habibovic et al., 2018; Marcus et al., 2024; Matthews et al., 2017). The strong preference for vehicle movement as a cue also supports existing literature on the significance of implicit communication in traffic interactions (Jayaraman et al., 2020; Zhao et al., 2022). However, the relatively low ranking of the audio speaker contrasts with some existing literature, which has suggested that auditory cues can enhance pedestrian understanding of ADS intentions (Marcus et al., 2024; Matthews et al., 2017).

When providing open-ended feedback on the eHMI, while some participants stated that the eHMI "design [was] clear and easy to understand," many noted that it was visually cluttered. Participants suggested improvements including universal symbols, flashing lights, larger fonts, and varied colors to enhance usability. They emphasized incorporating familiar traffic elements such as white pedestrian walking symbols, red hand symbols, and crosswalk tones.

Many participants highlighted the need for inclusive design to accommodate diverse pedestrian populations, including those who are colorblind, cannot read, have other disabilities, speak different languages, or face challenging weather conditions affecting visibility. These observations align with existing literature (Eisma et al., 2019; Deb et al., 2018; Dey et al., 2021). An interesting discrepancy emerged between quantitative and qualitative feedback: although the audio speaker received low necessity rankings in ratings, it was repeatedly emphasized in open-ended responses as important, especially for visually impaired or otherwise disabled pedestrians.

## DISCUSSION

**Impact of eHMI.**

In this study, with the eHMI equipped, pedestrians made more confident decisions around level-5 ADSs, crossing more and earlier, and felt safer and trusting in the vehicles. This aligns with results from previous studies that saw similar effects from the inclusion of an eHMI (Deb et al., 2018; Dey et al., 2021; Ferenchak & Shafique, 2022; Lagström & Lundgren, 2015; Matthews et al., 2017). This suggests that the eHMI acts as an effective form of explicit communication between the pedestrian and level-5 ADS, acting to fill the role of eye contact, hand gestures, and other forms of direct communication with drivers that help pedestrians understand if they have been detected and if it is safe to cross the road. And, the increased perceptions of safety, trust, and understanding afforded by the eHMI suggest that it made traffic interactions clearer and more efficient, helping pedestrians make informed road crossing decisions.

In the eHMI itself, this study found that visual features were preferred over audio cues, with the most preferred ones being a text-display, label, external speedometer, then pedestrian silhouette. This aligns with previous studies that similarly found text to be preferred by pedestrians for its clarity (Deb et al., 2018; Ferenchak & Shafique, 2022; Wang et al., 2021). The high ranking of the label also aligns with previous studies that found it important to inform pedestrians that the vehicle was automated (Habibovic et al., 2018; Rasouli & Tsotsos, 2020; Lagström & Lundgren, 2015; Wang et al., 2021). The external speedometer was likely ranked highly because it gave explicit information about implicit movement, directly quantifying changes in speed. The pedestrian silhouette was based on American traffic signals, in which a walking man is used at crosswalks to indicate when pedestrians can cross; other studies have suggested that the use of symbols to indicate level-5 ADS intent is helpful (Deb et al., 2018; Habibovic et al., 2018). However, in this study, the text- and symbol-based visual cues gave advice directly to pedestrians, which Habibovic et al. (2018) does not recommend, as the level-5 ADS may indicate its yielding intent with an advice based signal while unaware of some other condition that may make crossing unsafe; other studies have therefore suggested that level-5 ADSs ought to advise pedestrians of their future actions, as opposed to telling them what to do. In addition, audio cues have been found to be useful to pedestrians given their salience (Matthews et al., 2017), and open-ended feedback in this study reflected that. This eHMI was also busy, including many different features. While it is useful to include different modalities of information (Deb et al., 2018; Eisma et al., 2019), too much can cause signal blindness, leading pedestrians to ignore the display if there is too much too process, seen in open ended feedback and previous studies (Dey et al., 2021).

This study's main limitation was the use of text-based scenarios, which inherently provide an abstracted interaction experience that may have influenced eHMI feature rankings. Additionally, convenience sampling may have limited sample diversity. Future research should focus on refining eHMI features and exploring new scenarios, various pedestrian groups, and immersive technologies like virtual and augmented reality for simulating realistic pedestrian-ADS encounters.

## CONCLUSION

This study examined the impact of an eHMI on pedestrians' road-crossing decisions around level-5 ADSs. The eHMI significantly influenced crossing behavior, leading to more confident and earlier crossings, while also improving perceptions of safety, trust, and understanding. Regarding the eHMI, pedestrians overall preferred the eHMI to be equipped, with the most necessary elements being visual, including text displays, labels, speedometers, and pedestrian silhouettes. Audio cues, while ranked lowest, were still considered important in feedback. Practitioners should prioritize visual eHMI features that clearly communicate ADS automation status and speed while ensuring designs remain simple to avoid perceptual overload and include audio cues for accessibility.

## REFERENCES

Deb, S., Strawderman, L. J., & Carruth, D. W. (2018). Investigating pedestrian suggestions for external features on fully autonomous vehicles: A virtual reality experiment. *Transportation Research Part F: Traffic Psychology and Behaviour*, *59*, 135–149.

Dey, D., Matviienko, A., Berger, M., Pfleging, B., Martens, M., & Terken, J. (2021). Communicating the intention of an automated vehicle to pedestrians: The contributions of eHMI and vehicle behavior. *It - Information Technology*, *63*(2), 123–141.

Eisma, Y. B., van Bergen, S., ter Brake, S. M., Hensen, M. T. T., Tempelaar, W. J., & de Winter, J. C. F. (2019). External Human–Machine Interfaces: The Effect of Display Location on Crossing Intentions and Eye Movements. *Information*, *11*(1), 13.

Ferenchak, N. N., & Shafique, S. (2022). Pedestrians' Perceptions of Autonomous Vehicle External Human-Machine Interfaces. *ASCE-ASME J Risk and Uncert in Engrg Sys Part B Mech Engrg*, *8*(3), 034501.

Habibovic, A., Lundgren, V. M., Andersson, J., Klingegård, M., Lagström, T., Sirkka, A., Fagerlönn, J., Edgren, C., Fredriksson, R., Krupenia, S., Saluäär, D., & Larsson, P. (2018). Communicating Intent of Automated Vehicles to Pedestrians. *Frontiers in Psychology*, *9*, 1336.

Hafeez, F., Sheikh, U. U., Al-Shammari, S., Hamid, M., Khakwani, A. B. K., & Arfeen, Z. A. (2023). Comparative analysis of influencing factors on pedestrian road accidents. *Bulletin of Electrical Engineering and Informatics*, *12*(1), 257–267.

Jayaraman, S. K., Tilbury, D. M., Jessie Yang, X., Pradhan, A. K., & Robert, L. P. (2020). Analysis and Prediction of Pedestrian Crosswalk Behavior during Automated Vehicle Interactions. *2020 IEEE International Conference on Robotics and Automation (ICRA)*, 6426–6432.

Lagström, T., & Lundgren, V. M. (2015). An investigation of pedestrian-driver communication and development of a vehicle external interface. *Human Factors*.

Marcus, V., Muldoon, J., & Motamedi, S. (2024). Pedestrian interaction with automated driving systems: Acceptance model and design of External Communication Interface. *Lecture Notes in Computer Science*, 63–82.

Matthews, M., Chowdhary, G., & Kieson, E. (2017). *Intent Communication between Autonomous Vehicles and Pedestrians* (arXiv:1708.07123). arXiv.

Métayer, N., & Coeugnet, S. (2021). Improving the experience in the pedestrian's interaction with an autonomous vehicle: An ergonomic comparison of external HMI. *Applied Ergonomics*, *96*, 103478.

Rasouli, A., & Tsotsos, J. K. (2020). Autonomous Vehicles That Interact With Pedestrians: A Survey of Theory and Practice. *IEEE Transactions on Intelligent Transportation Systems*, *21*(3), 900–918.

Shi, E., Gasser, T. M., Seeck, A., & Auerswald, R. (2020). The Principles of Operation Framework: A Comprehensive Classification Concept for Automated Driving Functions. *SAE International Journal of Connected and Automated Vehicles*, *3*(1), 12-03-01–0003.

Wang, P., Motamedi, S., Qi, S., Zhou, X., Zhang, T., & Chan, C.-Y. (2021). Pedestrian interaction with automated vehicles at uncontrolled intersections. *Transportation Research Part F: Traffic Psychology and Behaviour*, *77*, 10–25.

WHO. (2023). Global status report on road safety 2023. World Health Organization. https://www.who.int/publications/i/item/9789240086517

Zhao, X., Li, X., Rakotonirainy, A., Bourgeois- Bougrine, S., & Delhomme, P. (2022). Predicting pedestrians' intention to cross the road in front of automated vehicles in risky situations. *Transportation Research Part F: Traffic Psychology and Behaviour*, *90*, 524–536.